\documentclass[%
 reprint,
superscriptaddress,
 bibnotes,
 amsmath,
 amssymb,
 aps
]{revtex4-2}

\usepackage{graphicx}
\usepackage{dcolumn}
\usepackage[title]{appendix}
\usepackage{bm}
\usepackage{braket}
\usepackage{xcolor}
\usepackage[caption=false]{subfig}
\usepackage[colorlinks=true,linkcolor=blue,citecolor=blue,filecolor=blue,urlcolor=blue]{hyperref}
\usepackage{lipsum} 
\definecolor{mygreen}{rgb}{0.2,0.7,0.2}  

\graphicspath{{Figures/}}
\begin{document}

\preprint{APS/123-QED}
\title{Excitons in nonlinear optical responses: shift current in MoS$_2$ and GeS monolayers}


\author{J.J. Esteve-Paredes}
\email{juan.esteve(at)uam.es}
\affiliation{Departamento de F\'isica de la Materia Condensada, Universidad Aut\'{o}noma de Madrid, E-28049 Madrid, Spain}
\author{M. A. Garc\'ia-Bl\'azquez}%
\affiliation{Departamento de F\'isica de la Materia Condensada, Universidad Aut\'{o}noma de Madrid, E-28049 Madrid, Spain}
\author{A. J. Ur\'ia-\'Alvarez}%
\affiliation{Departamento de F\'isica de la Materia Condensada, Universidad Aut\'{o}noma de Madrid, E-28049 Madrid, Spain}
\author{M. Camarasa-G\'omez}
\affiliation{Department of Molecular Chemistry and Materials Science, Weizmann Institute of Science, Rehovot 7610001, Israel}
\author{J.J. Palacios}
\affiliation{Departamento de F\'isica de la Materia Condensada, Universidad Aut\'{o}noma de Madrid, E-28049 Madrid, Spain}
\affiliation{Instituto Nicol\'as Cabrera (INC), Universidad Aut\'onoma de Madrid, E-28049 Madrid, Spain.}
\affiliation{Condensed Matter Physics Center (IFIMAC), Universidad Aut\'onoma de Madrid, E-28049 Madrid, Spain.}
\date{\today}

\begin{abstract}

It is well-known that exciton effects are determinant to understand the optical absorption spectrum of low-dimensional materials. However, the role of excitons in nonlinear optical responses 
has been much less investigated at an experimental
level. Additionally, computational methods to calculate nonlinear conductivities in real materials are still not widespread, particularly taking into account excitonic interactions. We present a methodology to calculate the excitonic second-order optical responses in 2D materials relying on: 
(i) ab initio tight-binding Hamiltonians obtained by Wannier interpolation
and (ii) the Bethe-Salpeter equation with effective electron-hole interactions. Here, in particular, we explore the role of excitons in the shift current of monolayer materials.  
Focusing on MoS$_2$ and GeS monolayer systems, our results show that $2p$-like excitons, which are dark in the linear response regime, yield a contribution to the photocurrent comparable to that of $1s$-like excitons. 
Under radiation with intensity $\sim 10^{4} $W/cm$^2$, the excitonic theory predicts 
in-gap photogalvanic currents of almost $\sim 10$ nA in sufficiently clean samples, which is typically one order of magnitude higher than the value predicted by independent-particle theory near the band edge.

\end{abstract}

\maketitle
\section{Introduction}
Excitons are key to our understanding of the linear optical response in low-dimensional semiconductors and insulators. Examples can be found in nanotubes \cite{chopra1995boron,wang2005optical} and quasi-two-dimensional (2D) crystals \cite{mak2010atomically,li2014measurement}. In theses systems, the combined effect of quantum confinement and a weak screening environment lead to large excitonic binding energies \cite{wirtz2006,ramasubramaniam2012large,qiu2013optical}. This fact manifests in strong peaks below the quasi-particle gap in photoluminiscence and optical absorption experiments. From the theory point of view, the linear response formalism plus the Bethe-Salpeter equation (BSE) for exciton calculations have been proven to be  powerful tools to explain the measured optical responses in the low-field regime
\cite{martin2016interacting, trolle2014theory,ridolfi2018excitonic,bieniek2020band,qiu2016screening,marsili2021spinorial,molina2016temperature}.

In the nonlinear regime, the bulk photovoltaic or photogalvanic effect \cite{sturman2021photovoltaic} (abbreviated BPVE or PGE) is a nonlinear effect of second order that has attracted much interest lately (see Ref. \cite{dai2023recent} for a review). It refers to the generation of a direct current (DC) in the bulk of a non-centrosymmetric material under light radiance, and thus it has an inherent potential for application in solar-cell devices. The BPVE was early explored in ferroelectric materials \cite{koch1975bulk, koch1976anomalous}, and continued gathering attention during the next decades \cite{von1981theory, batirov1997bulk, fridkin1993bulk, hornung1983band, buse1997light} until nowadays, as experiments achieve a high-efficiency in power conversion \cite{spanier2016power}. More recently, low-dimensional materials are also being explored in this regard. For instance, it has been shown that a large BPVE occurs in WS$_2$ nanotubes 
\cite{zhang2019enhanced,krishna2023understanding} and nanoribbons\cite{science.adn9476}, strained 3R-MoS$_2$ and phosporene \cite{dong2023giant,sun2024strain}, monolayer MoSe$_2$ \cite{quereda2018symmetry}, SnS \cite{chang2023shift,liang2023strong}, or one-dimensional grain boundaries \cite{zhou2024giant}.

The characterization and rationalization of the experimental signatures of the BPVE is a current active field of research\cite{sotome2019spectral,sotome2021terahertz}. On the theoretical side, methods based on the independent-particle approximation (IPA) for the band structure and Bloch eigenstates are widely used to calculate the second order and frequency-dependent conductivities
\cite{sipe2000second,young2012first,rangel2017large,cook2017design,tan2016shift,xiao2022non,sauer2023shift,zhang2019switchable,zhang2022tailoring,chaudhary2022,kaplan2022twisted}. This level of theory is sometimes successful, specially in bulk materials \cite{sotome2019spectral,young2012first,ni2021giant,puente2023abinitio}, 
while in other experiments the measured values of the photogalvanic currents are nearly one order of magnitude different than the predicted ones \cite{zhang2019enhanced,krishna2023understanding}, or show features that indicate the need for a many-body description of the problem
\cite{sotome2021terahertz}. It is therefore necessary to develop a more complete theory of the nonlinear optical response that goes beyond a purely IPA description, in order to achieve a reliable characterization of the BPVE in real materials. As shown below, including electron-hole interactions makes a huge difference.

In the literature, electron-hole interaction effects have been addressed in two scenarios. First, in the study of the so-called \textit{ballistic} current \cite{dai2023recent}, which originates from asymmetric carrier generation in the Brillouin zone (BZ) and is considered to be an 
incoherent process (with its own relaxation time).  Here,
electron-hole interactions seem to have a minor effect both in bulk and 2D systems \cite{dai2021first}. 
On the other hand, more recent works have considered the effect of excitons within the coherent nonlinear optical response formalism
\cite{huang2023large,konabe2021exciton}. Additionally, the role of excitons in 2D materials has been investigated by Green's function methods in the time domain \cite{chan2021giant,hu2023hu}. These few works show that excitons are equally or even more dominant in the non-linear response of low-dimensional materials than in the linear response. 

In this work, we present a methodology to compute the second-order optical conductivity which takes into account electron-hole interactions, updating previous approaches to linear optical properties based on tight-binding methods \cite{trolle2014theory,wu2015exciton, galvani2016excitons,ridolfi2018excitonic,bieniek2020band,uria2023efficient}. Here we start from an ab initio tight-binding description of the band structure, obtained by Wannier interpolation.  Importantly, the use of a Wannier Hamiltonian  ensures that all relevant dipole matrix elements in the local basis are properly accounted for, which gives a  description of optical responses more accurately than a purely effective tight-binding approximation  \cite{wang2017first,ibanez2018abinitio,ibanez2022assessing}.
Secondly, electron hole interactions are included by solving the BSE with effectively-screened electron-hole interactions. As a proof of concept, we focus on the evaluation of the shift current, which, in the absence of magnetism and under radiation with linearly-polarized light. This current-response is clearly discernible in experiments and understood with theoretical calculations \cite{sotome2019spectral, sotome2021terahertz,young2012first,koch1975bulk,koch1976anomalous,puente2023abinitio}.

As case examples, we study  $2H$-MoS$_2$ and  GeS monolayers. The former is a  prototypical TMDC material, showing a rich structure of exciton levels due to the important role played by spin-orbit coupling (SOC), while the latter is often taken as a test case for its simplicity and possibly very large  
shift current \cite{cook2017design,rangel2017large, ibanez2018abinitio,chan2021giant,chan2021giant}.
We compute the shift current as a function of the frequency of the incident radiation, as it would be measured in a photocurrent spectroscopy experiment \cite{quereda2018symmetry,zhang2019enhanced}.  We show that excitonic effects red-shift the IPA frequency-dependent response function, generally increasing the shift current by nearly an order of magnitude. This happens by the appearance of 
a finite and sometimes dominant contribution at energies below the band gap.
Interestingly, we also show that $2p$-like excitons, dark in linear response, are \textit{bright} in the second-order optical response and, therefore, can give a contribution to the BPVE.




Our work is organized as follows. In Sec. \ref{sec:ii} we review the theoretical framework to calculate exciton states and to  evaluate the second-rank conductivity tensor corresponding to the shift current. In Sec. \ref{sec:iii} we show our numerical results of the different frequency-dependent quantities, both with and without electron-hole interactions. Finally, in \ref{sec:iv} we summarize our conclusions and future-work perspectives.

\section{Methods}

\label{sec:ii}
\subsection{Quasi-particle band structure and excitons}
From a many-body perspective, the study of the exciton physics in a gapped material requires first the evaluation of its quasi-particle band structure. To carry out this task, one typically introduces electron-electron interactions beyond the Kohn-Sham density functional theory (DFT) level by calculating a self-energy correction to the bands, $\varepsilon_n(\bold{k})=\varepsilon_n^{\rm{KS}}(\bold{k})+\Sigma_n(\bold{k})$. The self-energy is typically obtained within the GW approximation \cite{martin2016interacting}, where dynamical and local-field effects are included. Alternatively, hybrid-functionals can be used to obtain a satisfactory quasi-particle band structure (for instance, the band gaps are larger than those obtained from standard DFT) at possibly lower computational cost than that of GW methods.

Within this scenario, we will assume that a Wannier interpolation \cite{mostofi2008wannier,pizzi2020wannier} to the best-available quasi-particle band structure has been carried out. We also assume that a set of well-localized $n_{\rm orb}$ Wannier orbitals, which we denote $w_\alpha(\bold{r}-\bold{d}_\alpha-\bold{R})=\braket{\bold{r}|\alpha \bold{R}}$, has been found. Bloch states built with these functions, 
$\ket{\alpha \bold{k}}=1/\sqrt{N}\sum_{\bold{R}}e^{i\bold{k}\cdot \bold{R}}\ket{\alpha \bold{R}}$, are referred to as Bloch-Wannier states.  The single-particle Hamiltonian projected on these states is given by
\begin{equation}
\label{eq:wannierhamiltonian}
H_{\alpha \alpha'}^{(\text{W})}(\bold{k})=\sum_{\bold{R}}e^{i\bold{k}\cdot \bold{R}}\braket{\alpha \bold{0}|\hat{H}|\alpha' \bold{R}}.
\end{equation}
Note from the phase factor in \eqref{eq:wannierhamiltonian} we choose the so-called \textit{lattice} gauge, whereas the \textit{atomic} gauge is another convention that is present in the literature \cite{vanderbilt2018berry, esteveparedes2023comprehensive} \footnote{See discussion around the \texttt{WannierTools} package: https://www.wanniertools.org/theory/tight-binding-model/ (accessed May 2024)}. Diagonalization of the $n_{\text{orb}}\times n_{\text{orb}}$ Hamiltonian matrix yields the Wannier-interpolated energy eigenvalues,
\begin{equation}
    \big[U^{\dagger}(\bold{k})H^{(\text{W})}(\bold{k})U(\bold{k})\big]_{nn'}=\varepsilon_n(\bold{k})\delta_{nn'},
\end{equation}
where $U(\bold{k})$ is the unitary transformation matrix containing the Bloch eigenstates coefficients $c_{n\alpha}(\bold{k})$ in the expansion of the band-structure eigenstates $\ket{n\bold{k}}=\sum_{\alpha}c_{n\alpha}(\bold{k})\ket{\alpha \bold{k}}$.

Excitons arise from the Coulomb interaction of a quasi-electron and quasi-hole, some of them forming bound states within the bandgap. Using valence ($v$) and conduction ($c$) single-particle states, we expand exciton states 
of zero center of mass momentum ($\bold{Q}=0$) in a basis of free electron-hole pairs as $\ket{X_N}=\sum_{vc\bold{k}}A_{vc\bold{k}}^{(N)}(\bold{k})\hat{c}_{c\bold{k}}^{\dagger}\hat{c}_{v\bold{k}}\ket{0}$, being $\ket{0}$ the ground state, identified as the Fermi sea. Exciton states are thus found by solving the BSE \cite{rohlfing2000electron} in the Tamm-Dancoff approximation
\begin{equation}
\label{eq:bse}
\begin{split}
     &(\varepsilon_{c\bold{k}} - \varepsilon_{v\bold{k}}) A_{vc}^{(N)}(\bold{k}) \\    & \ \ \ \ +\sum_{v'c'\bold{k}'}\braket{vc\bold{k}|K_{\text{eh}}|v'c'\bold{k}'}A_{v'c'}(\bold{k}')=E_N A_{vc}(\bold{k}),
\end{split}
\end{equation}

where $K_{\text{eh}}=-(D-X)$ is the electron-hole interaction kernel, including the direct ($D$) and exchange terms ($X$). The direct term reads
\begin{equation}
\label{eq:directcoulomb}
    D_{vc,v'c'}(\bold{k},\bold{k}')=\int \psi_{c\bold{k}}^{\ast}(\bold{r}) \psi_{v'\bold{k}'}^{\ast}(\bold{r}') W(\bold{r},\bold{r}')\psi_{c'\bold{k}'}(\bold{r}) \psi_{v\bold{k}}(\bold{r}'),
\end{equation}
and we treat the screened Coulomb interaction $W$ in the static limit. The exchange term in the kernel is obtained by interchanging $c'\bold{k}'$ and $v\bold{k}$. 
Fully ab initio calculations, namely GW-BSE methods, often employ an unscreened exchange interaction, as derived using many-body perturbation theory (MBPT) techniques \cite{rohlfing2000electron}.
Recent studies point out the necessity of constructing an effective screening for the exchange in some scenarios \cite{qiu2023hilbert}, such as core-conduction excitations. In our variational approach where the BSE can be seen as the result of representing the Hamiltonian in a Hilbert space expanded by one-electron excitations Slater determinants, it is natural to assume the same screening for all Coulomb matrix elements \cite{bieniek2020band, bieniek2022theory}.

In this work, we model the 2D screening environment by using a Rytova-Keldysh potential \cite{cudazzo2011dielectric}. In the case of vacuum surroundings, this model has one parameter, the screening length $r_0$, which can actually be calculated from DFT \cite{cudazzo2011dielectric}. The choice of the 2D interaction model may be determinant for the resulting exciton energy series and the envelope wave functions (recent computational packages \cite{uria2023efficient,dias2023wantibexos} allow to explore other different potentials). In the case of layered materials of much interest such as heterostructures or one-dimensional systems, other interaction models have also been proposed \cite{kamban2020interlayer, villegas2024screened} and can be considered for future studies.

The Wannier orbitals $w_\alpha(\bold{r}-\bold{d}_\alpha-\bold{R})=\braket{\bold{r}|\alpha \bold{R}}$  have a finite spread that may be of the order of 
several unit cells. To efficiently implement Eq. \eqref{eq:directcoulomb}, here we ignore the real-space details of the orbitals, treating them as delta-like functions peaked at their centers \cite{ridolfi2018excitonic,uria2023efficient,dias2023wantibexos,wu2015exciton}. The actual shape of the orbitals has been taken into consideration by other authors \cite{bieniek2022theory}, although we skip this numerical refinement in exchange for less computational and technical burden. The computational efficiency of this approach depends significantly on whether Coulomb integrals are performed in real or reciprocal space. We use a real-space implementation \cite{wu2015exciton}, which allows to cast the Coulomb interaction matrix elements in a billinear form that is amenable for vectorizing calculations \cite{uria2023efficient}. Below, we show that our results for the exciton structure in monolayer MoS$_2$ are in agreement with previous literature, which is a successful test for our methodology.

\subsection{Nonlinear optical response}
In a general manner, the frequency-dependent second-order current-density reads 
\begin{equation}
\begin{split}
    j_a ^{(2)} (\omega)&= \int d\omega_p \int d\omega_q \sum_{bc} \sigma_{abc}^{(2)}(\omega;\omega_p,\omega_q) \\
    & \times \varepsilon_{b}(\omega_p)\varepsilon_{c}(\omega_q)  \delta([\omega_p+\omega_q]-\omega)
\end{split}
\end{equation}
The contribution to the BPVE can be extracted by taking the DC limit. Here, we focus on the response under radiation with linearly-polarized light, known as shift current (in the absence of magnetic effects). The amount and directionality of the DC shift-current response depends on the frequency of the incident pulse, which can can be expressed in terms of the second-order conductivity tensor at zero frequency \cite{sipe2000second,ahn2020low,garcia2023shift}:
\begin{equation}\label{eq_shiftCurrent}
 j_a^{(2,\text{sh})}=2\sum_{bc}\text{Re}[\sigma_{abc}^{(\text{sh})}(0,\omega,-\omega)\varepsilon_b(\omega)\varepsilon_c(-\omega)],
\end{equation}
where we have considered radiation under monochromatic light. The shift conductivity tensor (or simply shift conductivity) can be derived either in the velocity or length light-matter interaction gauges. Only the latter choice ensures the fastest convergence rates as well as the most accurate spectra for any basis size, and is also free from low-frequency divergences (see Ref. \cite{taghizadeh2017linear} and references therein). In the IPA,  the shift conductivity tensor reads

\begin{equation}
\label{eq:shift1}
\begin{split}
\sigma_{abc}^{(\text{sh})}(0,\omega,-\omega)=&-\frac{i\pi e^3}{2\hbar V}\sum_{nn' \bold{k}}
f_{nn' \bold{k}} [I_{nn' \bold{k}}^{abc}+I_{nn' \bold{k}}^{acb}] \\
& \times \delta(\omega-\omega_{nn' \bold{k}}),
\end{split}
\end{equation}
where the strength of the transitions is $I_{nn' \bold{k}}^{abc}=r_{nn' \bold{k}}^{b}r_{n'n \bold{k}}^{c;a}$, and the matrix elements are
\begin{equation}
\begin{split}
&r^{a}_{nn' \bold{k}}=(1-\delta_{nn'})\xi^{a}_{nn'\bold{k}}, \\
&r_{nn' \bold{k}}^{a;b}=\nabla_{b}r_{nn' \bold{k}}^{a}-ir_{nn' 
\bold{k}}^{a}[\xi^{b}_{nn\bold{k}}-\xi^{b}_{n'n'\bold{k}}],
\end{split}
\end{equation}
being $\xi^{a}_{nn'\bold{k}}$ the Berry connection and $r_{nn' \bold{k}}^{a;b}$ the generalized derivative of $r^{a}_{nn' \bold{k}}$. While the off-diagonal Berry connections can be related to matrix elements of velocity operator, the generalized derivative presents the issue of computing the diagonal elements $\xi^{a}_{nn\bold{k}}$, which suffer from phase indetermination in numerical approaches. To circumvent this problem, an alternative approach is the use of sum rules \cite{taghizadeh2017linear} to directly evaluate $r_{nn' \bold{k}}^{a;b}$, now only involving velocity matrix elements between remote energy bands. In a Wannier-interpolation scheme, these remote bands may be out of the wannierized subspace and therefore not available to be included in calculations. While this problem can still be mitigated by means of perturbation theory \cite{ibanez2018abinitio}, here we set a numerical algorithm to smooth the phase of eigenvectors (see our previous work \cite{garcia2023shift} and the {\textit{Supplementary Material}). As a result, $\bold{k}$-derivatives of matrix elements can be evaluated numerically and the generalized derivative can be computed \textit{exactly}. 

In a similar fashion, a version of Eq. \eqref{eq:shift1} including exciton effects can be written \cite{pedersen2015intraband,taghizadeh2018gauge}, but now involving matrix elements between excitonic many-body states. We extract the shift conductivity by taking the real part of the second order conductivity tensor at zero frequency \cite{taghizadeh2018gauge}

\begin{equation} 
\label{eq:shiftexcitonfinal}
\begin{split}
& \sigma_{abc}^{(\text{sh})}(0;\omega,-\omega)= \\
& \ \ \ \frac{e^3}{\hbar V}\sum_{NN'} \text{Re}(S_{NN'}^{abc}) \times[\delta(\hbar \omega-E_N)+\delta(\hbar \omega+E_N)].
\end{split}
\end{equation}
where we have defined the exciton coupling matrix element $S_{NN'}^{abc}\equiv iR_{0N}^aR_{NN'}^bR_{N'0}^c$, being $R_{0N}^a=\braket{\rm{GS}|\hat{r}_a|X_N}$ and $R_{NN'}^a=\braket{X_N|\hat{r}_a|X_{N'}}$ with GS denoting the ground state. Further details about excitonic matrix elements are given in the \textit{Supplementary Material}.
The same phase criteria is applied to single-particle eigenvectors when constructing the BSE matrix in Eq. \eqref{eq:bse} (note that it depends on the single-particle eigenvectors via the direct and exchange terms), which allows to agree on the phase choice between $A_{vc\bold{k}}^{(N)}$ and single-particle matrix elements, ensuring that exciton matrix elements are properly evaluated.

\subsection{Post-processing workflow}
We have developed a post-processing workflow to evaluate Eqs. \eqref{eq:shift1} and \eqref{eq:shiftexcitonfinal}. 
After the evaluation of the DFT band structure, the \textsc{Wannier90} \cite{pizzi2020wannier} code is used to obtain a 
maximally-localized Wannier interpolation of the quasi-particle band structure. The Hamiltonian and dipole matrix elements between Wannier orbitals are printed, which is sufficient to evaluate the linear and nonlinear responses in the IPA. To include exciton effects, the output of \textsc{Wannier90} is interfaced with \textsc{Xatu} \cite{uria2023efficient}, which yields the exciton energies and wavefunctions as output. Eq. \eqref{eq:shiftexcitonfinal} is finally implemented using the previous building blocks. 
\section{Numerical results} 
\label{sec:iii}
\subsection{1L-MoS$_2$}
\subsubsection{Band structure and computational details}
\label{sec:iiia}
We start by calculating the quasiparticle band structure of monolayer MoS$_2$, as shown in Fig. \ref{fig:mos2_bands}(a). In the single-particle calculations, a SRSH functional \cite{ramasubramaniam2019transferable,camarasagomez2023transferable} has been employed, yielding a direct gap of 2.60 eV at the $K$ point. We include spin-orbit coupling  (SOC) in the calculation, yielding a top valence band splitting of 0.17 eV. These values are in good agreement with previous calculations 
(see e.g. \cite{qiu2016screening,marsili2021spinorial}) . In a post-processing step, we obtain a Wannier Hamiltonian \cite{mostofi2008wannier,pizzi2020wannier} for the manifold of the 16 highest valence bands and 8 bottom conduction bands.

Next, we include exciton effects by solving the BSE. We use our recently developed package \textsc{Xatu} to solve Eq. \eqref{eq:bse}, as explained previously. To extract our results, we perform calculations using a dimension of $N_k^2\times n_v \times n_c=102^2\times 2 \times 2$ in the Hilbert space when solving the BSE. For further analysis, we name the energy bands as $(v_{-1},v_0,c_1,c_2)$ in order of increasing energy at the $K$ point. We consider a screening length of $r_0=35$ \AA \  and vaccuum suroundings in the Rytova-Keldysh potential, similarly to previous studies \cite{berkelbach2013theory}. For a fair comparison between the non-interacting and excitonic results, below we restrict to the same subspace of bands when evaluating the optical responses in the IPA. 

\begin{figure}[h!]
   \centering
   \includegraphics[width=8.5cm]{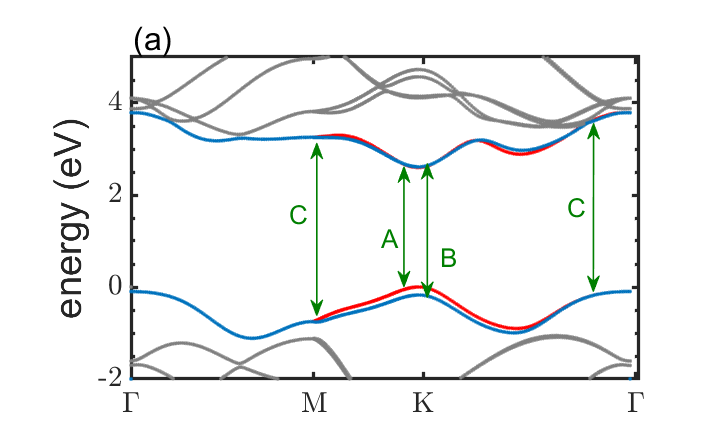}
    \includegraphics[width=8.5cm]{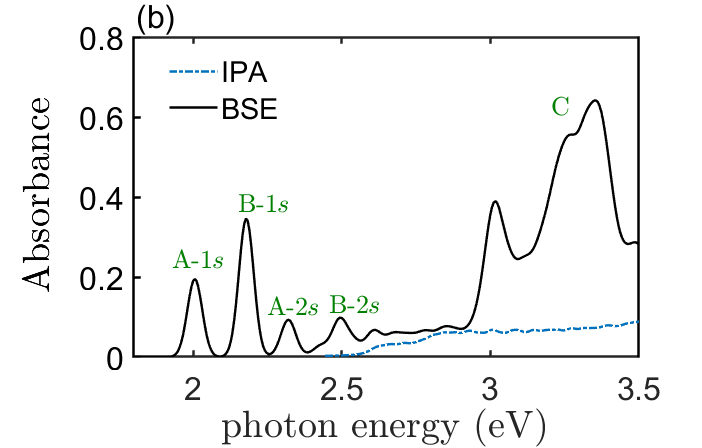}

   \caption{(a) Band structure of monolayer MoS$_2$ including SOC. Blue (red) bands have its eigenstates formed by mostly spin-up (spin-down) Bloch-Wannier states. We have pointed the position of the most relevant electron-hole excitation to the wavefunction of the labeled excitons in the absorbance spectrum. Bands that are not included in the BSE calculation are in grey color. (b) Absorbance of monolayer MoS$_2$ as calculated in the independent-particle approximation and including exciton effects.}
   \label{fig:mos2_bands}
\end{figure}

\subsubsection{Exciton spectrum}
We briefly review here the characteristics of the excitons in MoS$_2$ based on the output of our methodology. To understand the exciton spectrum, special attention needs to be paid to the spin structure of the BSE. Since $S_z$ is a good quantum number in the vicinity of the $K$ and $K'$ points, the interacting BSE Hamiltonian almost decouples into a set of solutions with like or unlike spin of electron-hole pairs. Only excitons formed by spin-conserving electron-hole pairs are expected to have enough oscillator strength to be bright. SOC-induced band-splitting gives rise to two series of excitons. Around the $K$ point, the lowest-energy electron-hole pair excitation with the same spin is $(v_{0},c_{1})$. Accordingly, equal-energy electron-hole excitations occur at $K'$ with inverted spin character, due to time reversal symmetry. Excitons that are mainly formed by these electron-hole free pairs constitute the A series. The next like-spin free electron-hole excitations, $(v_{-1},c_{2})$ gives a different series of excitons, called B.  

Within the A series, we find the following structure for the exciton shell structure: $1s$ at 2.00 eV, $2p$ at 2.26 eV, $2s$ at 2.32 eV, $3d$ at 2.37 eV, and $3s$ at 2.43 eV. A similar B series is found blueshifted (in overall) by 0.17 eV, which is the $A$-$B$ split of the 1s exciton. The angular momenta character is asigned by inspecting the node structure of $A_{vc}^{(N)}(\bold{k})$. Our numerical results show that the lowest-energy exciton of the spectrum is formed by unlike-spin transitions around $K$ and $K$' valleys, redshifted from the optically bright A-$1s$ exciton by 25 meV. The previous spin structure of levels and our numerical results are in very good agreement with ab initio, tight-binding and model calculations \cite{qiu2013optical,qiu2015nonanalycity,qiu2016screening, wu2015exciton,ridolfi2018excitonic,bieniek2020band}, which tests our methodology and allow us to step into the study of optical responses.


\subsubsection{Light absorbance}
We start exploring the optical responses by examining the frequency-dependent absorbance of monolayer MoS$_2$, which is based on the evaluation of the linear conductivity (see the \textit{Supplementary Material}). For the current analysis, we have calculated the frequency-dependent spectrum both in the IPA approximation and after including excitons. We set a broadening $\eta=25$ meV throughout the rest of the work.

In Fig. \ref{fig:mos2_bands}(b) we show the frequency-dependent absorbance of light under linearly polarized light. The IPA curve displays a typical step-like shape up to our range of inspection, as more bands come into play in the M$-K-\Gamma$ path. 
When interactions are considered, the bright excitons become visible below the quasi-particle band gap. As discussed, SOC induces two series of excitons which are clearly identified. 
Higher in energy, we identify the noticeable C resonance around 3 eV, which corresponds to the excitons formed by electron-hole pairs in the continuum. Previous studies have shown that more bands are needed in order to fully resolve the magnitude of the C resonance \cite{ridolfi2018excitonic}, hence we do not analyze it here any further.
Apart from bright $s$-type excitons, the $p$-like ones can also be optically active in $D_{3h}$ monolayers \cite{galvani2016excitons,zhang2022intervalley}. However, due to the small trigonal warping effect in MoS$_2$ around the $K$ point, the latter have a vanishing oscillator strength \cite{quintela2023anisotropic} and are not visible in absorbance or transmittance spectra, in agreement with our numerical calculations.

\subsubsection{Shift conductivity}
We now turn into the main set of results in this work, in regards to the nonlinear optical response. We start by evaluating the frequency-dependent shift conductivity tensor entries, as explained in the theory section. If the $x$ axis is along the armchair direction, and taking into account the $D_{3h}$ symmetry of monolayer TMDCs, one can check that $\sigma_{xxx}=-\sigma_{xyy}=-\sigma_{xyx}=-\sigma_{xxy}$  while the other components are null\cite{garcia2023shift}. 
Note that the symmetry arguments to determine the shape of the tensor are the same as in the single-particle picture, since they can be made from Eq. \eqref{eq_shiftCurrent}. We therefore restrict our analysis to $\sigma_{xxx}$. 
Below, we present results neglecting the exchange interaction in the BSE, which is a common assumption in the literature when using effective models. Actually, the inclusion of the exchange interaction noticeably changes the results, as shown in the \textit{Supplementary Material}, although our  discussion is equally valid with and without exchange effects.

In Fig. \ref{fig:mos2_shift}, we show $\sigma_{xxx}^{(\text{sh})}(0;\omega,-\omega)$ for monolayer MoS$_2$, as computed in the IPA using Eq. \eqref{eq:shift1} and including exciton effects using Eq. \eqref{eq:shiftexcitonfinal}. 
The IPA shift conductivity is positive and barely varying in the shown energy range.
Interactions red-shift features that can be found in the IPA calculations at larger energies (see \textit{Supplementary Material} for the complete IPA curve), bringing them into the selected energy window, while adding new ones. First, we identify the in-gap exciton resonances and label its angular momentum character. We observe that the A-$1s$ and B-$1s$ resonances have a response of $\sim 8$ $\text{nm} \cdot \mu A/V^2$, which is roughly 3 times bigger than the IPA response near the band-gap. Differently from the linear absorbance case, we observe peaks corresponding to $p$ excitons. This feature can be understood by comparing the expressions of the linear and nonlinear oscillator strengths, and the magnitude of dipole matrix elements between exciton states. In MoS$_2$, the low trigonal warping around the $K$ points yields that $|R_{0,2p}| \simeq 0.05|R_{0,1s}|$, and as the linear oscillator strength is  $\sim |R_{0N}|^2$, the brightness of $p$ excitons is strongly supressed with respect to $s$ excitons. In contrast, the nonlinear oscillator strength behaves as $S_{NN'}\sim R_{0N}R_{NN'}R_{N'0}$ (normalized to the total volume). The similar value of the conductivity for the $1s$ and $2p$ excitons is understood by inspecting their coupling matrix elements
Our results yield that $|R_{1s,1s}|$ and $|R_{2s,2p}|$.  are particularly small, so inter-exciton couplings are responsible for the finite shift conductivity at those resonances. Our numerical results show that $|R_{1s,2p}| \sim 10 |R_{0,1s}| \sim 200|R_{0,2p}|$. The small value of $R_{0,2p}$ in the triple product of matrix elements is counteracted by a strong inter-exciton coupling and allows the $p$ exciton to be visible in the shift conductivity spectrum  with a similar brightness than the $1s$ exciton. This selection rules have been treated previously in the second harmonic generation optical response \cite{taghizadeh2019nonlinear}, and also by recent ab initio calculations \cite{ruan2023excitonic}.

Above the quasi-particle gap, we observe the C resonance whose shape is similar to the non-interacting case. Previous results based on tight-binding excitons \cite{ridolfi2018excitonic} have shown than adding more bands to the calculation can enlarge the optical response of this part of the spectrum up to $25$ \%. Calculations with larger Hilbert spaces, needed to converge this part of the spectrum, are postponed for future work.
\begin{figure}[h!]
   \centering
   \includegraphics[width=8.5cm]{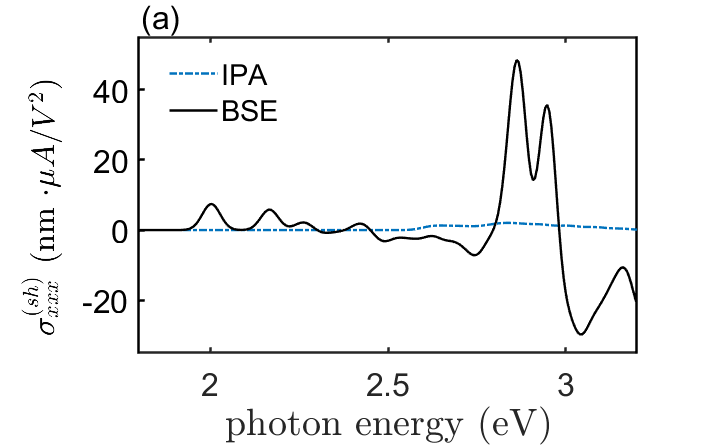}
   \includegraphics[width=8.5cm]{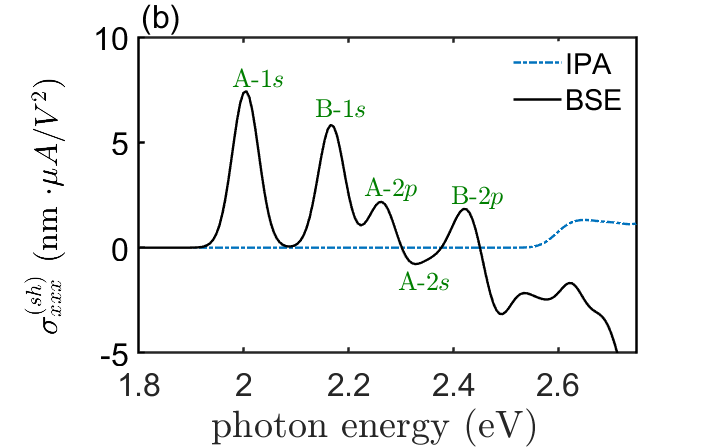}
   \caption {(a) Frequency-dependent shift conductivity of monolayer MoS$_2$ calculated in the IPA approximation and including exciton effects. See main text for discussion. (b) In-gap excitonic shift conductivity, as zoomed in from (a). Resonances are labeled according to the exciton spectrum.} 
   \label{fig:mos2_shift}
\end{figure}




\subsection{1L-GeS}
\subsubsection{Band structure and exciton spectrum}
We now revisit the case of monolayer GeS, whose shift current response in presence of exciton effects has been investigated by an ab initio time-dependent scheme \cite{chan2021giant}. Our analysis below adds extra insightful information about the relevant excitons giving rise to a large shift conductivity.

In Fig. \ref{fig:GeS_bands}(a) we show the band structure of monolayer GeS calculated using a SRSH functional. SOC is not included as it is expected to be small. Our calculations yield a direct gap at $\Gamma$ of 2.72 eV. In a post-processing step, we obtain a Wannier Hamiltonian \cite{mostofi2008wannier,pizzi2020wannier} for the manifold of the top 20 valence bands and first 7 conduction bands. Next, we include exciton effects by solving the BSE using the same dimension than in the previous example. In this case, we set a screening length to $r_0=20$ \AA \cite{gomes2016strongly}.  

According to Fig. \ref{fig:GeS_bands}(a), the lowest-energy and direct electron-hole excitation across the bandgap occurs at $\Gamma$, followed by a excitation along the $\Gamma$-X path, which we denote with $V_x$ following a previous study \cite{gomes2016strongly}. As a consequence, our calculations show that the first five bound exciton states are formed by a linear combination of $e$-$h$ pairs around $\Gamma$, while the fifth in energy is formed in the $V_x$ region. We label them according to its angular momentum character: $\Gamma$-$1s$ at 1.82 eV, $\Gamma$-$2p_1$ at 2.11 eV, $\Gamma$-$2p_2$ at 2.21 eV, $\Gamma$-$3p$ at 2.26 eV, and $V_x$-$1s$ at 2.36 eV. We elaborate in the following on the optical response of such excitons in the optical response.
 \begin{figure}[h!]
   \centering
   \includegraphics[width=8.5cm]{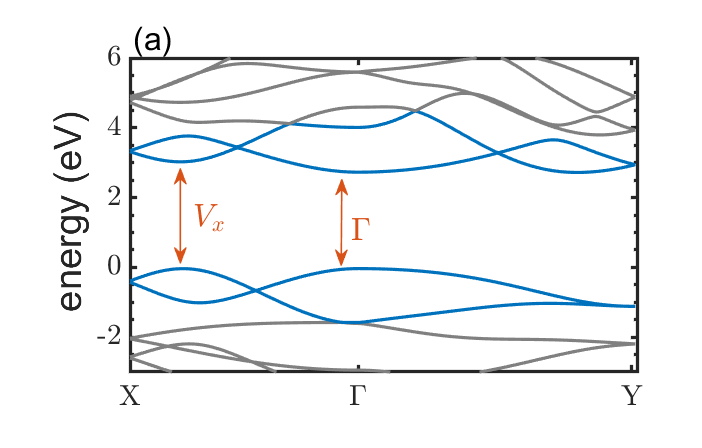}
   \includegraphics[width=8.5cm]{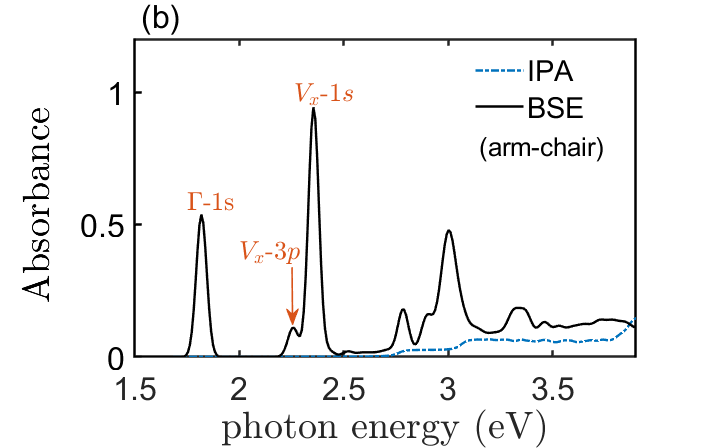}
   \includegraphics[width=8.5cm]{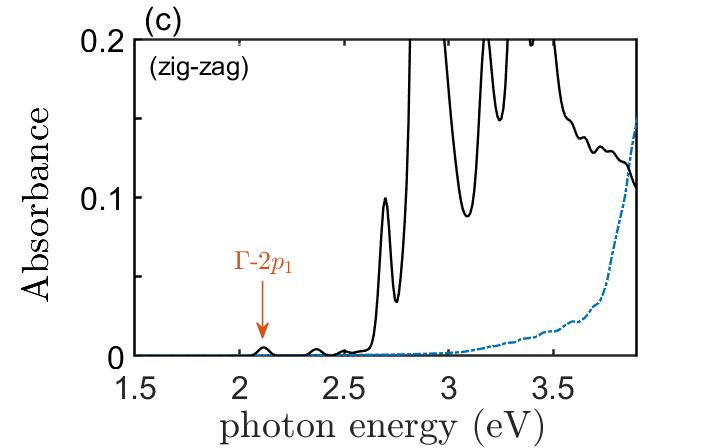}
   \caption{(a) Band structure of monolayer GeS (SOC has been not included).  Arrows mark the position of the most relevant electron-hole excitation to the wavefunction of the labeled excitons in the absorbance spectrum. Bands that are not included in the BSE calculation are in grey color. (b) Absorbance of monolayer GeS as calculated in the IPA approximation and including exciton effects for light polarized in the arm-chair direction. (c) Same as (b) but for zig-zag polarization.}
   \label{fig:GeS_bands}
\end{figure}

 \begin{figure}[h!]
   \centering
   \includegraphics[width=8.5cm]{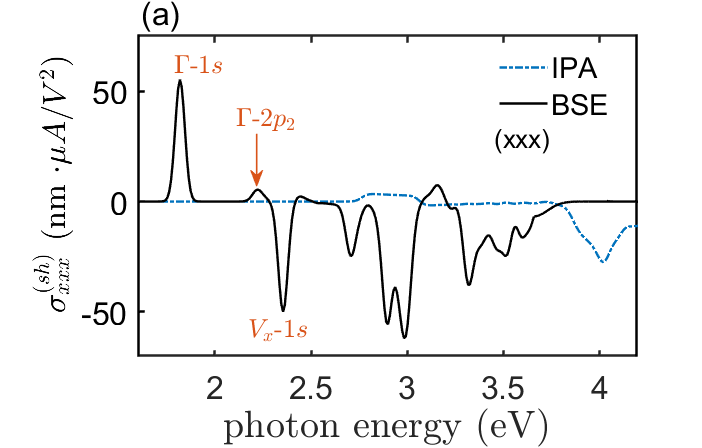}
   \includegraphics[width=8.5cm]{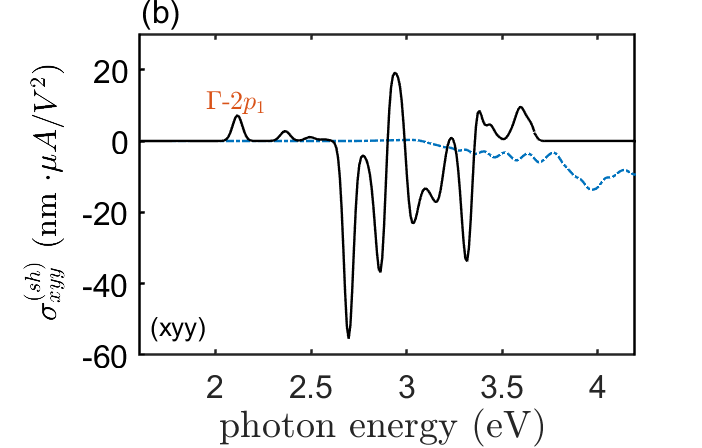}
   \caption {(a) Component $xxx$ of the frequency-dependent shift conductivity tensor of monolayer GeS calculated in the IPA approximation and including exciton effects. (b) Same for the $xyy$ component.} 
   \label{fig:GeS_shift}
\end{figure}
\subsubsection{Light absorbance and shift conductivity}

Fig. \ref{fig:GeS_bands}(b) shows the optical absorbance with and without exciton effects, using light polarized along the armchair (aligned with the $x$ axis) and zigzag directions of monolayer GeS. The case of armchair polarization shows two big peaks that are assigned to the $\Gamma$-$1s$ and $V_x$-$1s$ excitons. Remarkably, the $\Gamma$-$3p$ exciton is also considerably bright, which denotes the departure from the isotropic and hydrogen-like model for excitons. The zigzag case, shown in Fig \ref{fig:GeS_bands}(c), shows an almost dark in-gap spectrum. We identify the first peak with the $\Gamma$-$2p_1$, whose oscillator strength is appreciable in this case. The $\Gamma$-$2p_2$ level is thus the only exciton that is dark in absorbance.

Due the $C_{2v}$ point group of monolayer GeS, in this case with the $c_{2}$ rotation axis along the $x$ (in-plane) direction, the non-vanishing in-plane components of the shift conductivity are $\sigma_{xxx}$, $\sigma_{xyy}$, $\sigma_{yyx}=\sigma_{yxy}$ \cite{garcia2023shift}. We focus our analysis in $\sigma_{xxx}$ and $\sigma_{xyy}$}, which are shown in Fig \ref{fig:GeS_shift}(a) and (b). 
The $\sigma_{xxx}^{(\text{sh})}$ component shows that its main in-gap contribution comes from $1s$-like excitons, with opposite directionality. This fact can be understood by considering the non-interacting limit for the shift conductivity: the $\Gamma$ and $V_x$ regions in the BZ give a contribution of different sign to the shift conductivity, which can be seen in the IPA curve around 3 eV. Thus, exciton formed in such regions inherit this feature. By inspecting excitonic matrix elements, we
conclude that the $\Gamma$-$1s$ exciton has main contributions from the intra-exciton $S_{\Gamma\text{-}1s,\Gamma\text{-}1s}$
and inter-exciton 
$S_{\Gamma\text{-}2p_2,\Gamma\text{-}1s}$.  The former matrix element is two orders of magnitude than the equivalent in Mo$S_2$. We attribute this difference to a larger effective mass occuring in monolayer GeS at $\Gamma$ plus the large anisotropy effect, which depart excitons in GeS from the hydrogen-like picture in the electron hole interaction.

As in MoS$_2$, the $\Gamma$-$p_2$ becomes bright in the shift conductivity spectrum. Our numerical results show that half of its response comes from the coupling 
$S_{\Gamma\text{-}1s,\Gamma\text{-}2p_2}$
while the other half comes from the coupling with excitons at higher energies. Finally, we check that the $V_x$-$1s$ exciton response arise mainly from its intra-exciton  coupling matrix elements. As shown in the \textit{Supplementary Material}, its wavefunction does not overlap with excitons formed around $\Gamma$, therefore the coupling with exciton at lower energies is vanishing. On the other hand, $\sigma_{xyy}^{(\text{sh})}$ shows little low-energy response, as no bright excitons can be coupled through the nonlinear oscillator strength. At higher energies, this tensor components reaches similar minimum values as the $xxx$ component.

\subsection{Shift current in short circuit}

If light is linearly polarized along the $x$ direction, the total photogalvanic (shift) current generated in a portion of an hexagonal 2D material with $D_{3h}$ point group ($c_{2}$ rotation axis along $x$) of thickness $d$ and width $w$ reads \cite{krishna2023understanding,fei2020shift}
\begin{equation}
\label{eq:totalcurrent}
J_{x}^{(2, \text{DC})}(\omega)=\eta_{\text{inc}}(\omega)\cdot\eta_{\text{abs}}(\omega) \cdot G_{xxx}(\omega) \cdot w \cdot I_x(\omega),
\end{equation}
where $\eta_{\text{inc}}(\omega)=1-R(\omega)$, 
$\eta_{\text{abs}}(\omega)=(1-e^{-\alpha(\omega)\cdot d})$ and $G_{xxx}(\omega)=2\sigma_{xxx}^{(\text{sh})}(\omega)[c\epsilon_0 \alpha(\omega)]^{-1}$,  known as the \textit{Glass coefficient}.
$R$ and $\alpha$ stand for reflection and attenuation (or absorption) coefficients, respectively. We omit the Cartesian components of such quantities due to the isotropic in-plane dielectric tensor (in the case of 
hexagonal 2D materials). It has been argued \cite{fei2020shift} that the total current is minored if exciton effects are taken into account in calculations, as the magnitude of $R$ and $\alpha$ increases, which translates in less current in Eq. \eqref{eq:totalcurrent}. However, the change in $\sigma_{xxx}$ after including exciton effects was not accounted for, which is crucial based on our previous analysis. In monolayer materials, $\alpha(\omega)^{-1}$ is typically several orders of magnitude smaller than $d$. Therefore $(1-e^{-\alpha(\omega)\cdot d}) \simeq \alpha(\omega)\cdot d$ 
and, in the limit of a 2D sample \footnote{We numerically check that Eqs. \eqref{eq:totalcurrent} and \eqref{eq:totalcurrent2d} give nearly identical curves (not shown).}, leads to
\begin{equation}
\label{eq:totalcurrent2d}
J_{x}^{(2, \text{DC})}(\omega)=\frac{\eta_{\text{inc}}(\omega)}{c\epsilon_0} \cdot 2\sigma_{xxx}(\omega)  \cdot I_x(\omega) \cdot d \cdot w ,
\end{equation}
which means
that the exciton influence is ruled by the changes of the shift conductivity and the reflectivity factor. We assume an intensity of $1.6\times10^{4}$ W/cm$^2$ and a sample width of 2 $\mu$m, as in experiment \cite{zhang2019enhanced,krishna2023understanding}. We take a monolayer effective thickness for MoS$_2$ as reported in Ref. \cite{liu2020temperature}, while we use the same value for GeS. In the \textit{Supplementary Material}, we also compare our results for the reflectivity ratio with those obtained after reading the refractive index of Ref. \cite{liu2020temperature}, showing a good agreement between the theoretical and experimental excitonic signatures in the spectrum. 

\begin{figure}
   \label{fig:totalshift}
   \centering
   \includegraphics[width=8.5cm]{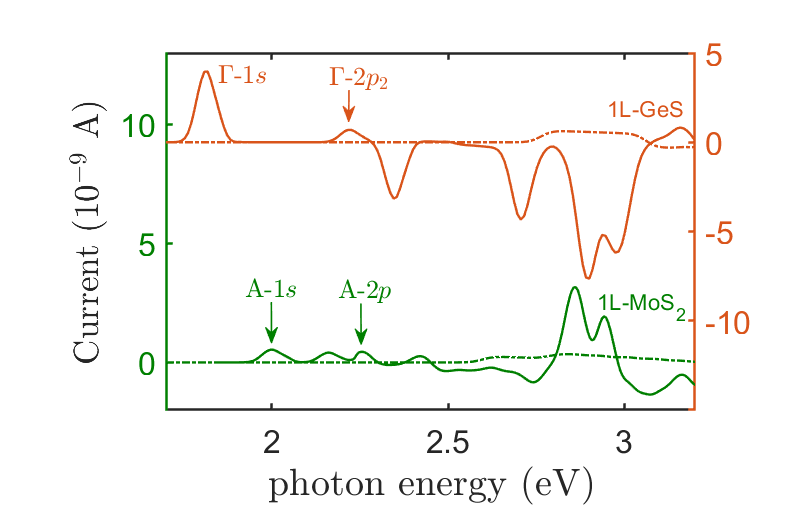}
   \caption{Total shift current as occurring in a short circuit configuration with 1L-MoS$_2$ and 1L-GeS$_2$, under radiation with monochromatic light. We show the calculation with (solid line) and without (dashed) exciton effects.}
\end{figure}

In Fig. \ref{fig:totalshift} we show the total current generated in a short-circuit configuration with samples of MoS$_2$ and GeS. We observe that in-gap $1s$ excitons give a response of $\sim 0.5\cdot10^{-9}$ nA and $\sim 5\cdot10^{-9}$ in MoS$_2$ and GeS, respectively; which corresponds to an enlargement by a factor of $\sim 2$ and $\sim 9$ with respect to the band-edge response predicted by the IPA calculation of each case. In general, the enhancement due to $e$-h interactions in the total current is not as large as in the shift conductivity [see Figs. \ref{fig:mos2_bands} and \ref{fig:GeS_shift}], due to an additional enhancement of the reflectivity ratio, i.e., $\eta_{\text{inc}}(\omega)$ is reduced in Eq. \eqref{eq:totalcurrent} and compensates changes in $\sigma_{xxx}^{(\text{sh})}$. Remarkably, the $2p$ resonances in monolayer MoS$_2$ have a similar magnitude than the $1s$. This is a consequence of the low oscillator strength of $2p$ excitons, which gives low reflection at such energies. This feature is not as visible in the $\Gamma$-2$p_2$ resonance of GeS, since an optically bright exciton is close in energy and gives a considerable reflectivity. 
At energies in the continuum, near 3 eV, the response is enlarged to $\sim 3$ nA in MoS$_2$, while it is inverted in GeS reaching $\sim -7$ nA, although more bands are needed to fully converge this part of the spectrum.

Photocurrent spectroscopy measurements in monolayer TMDCs allow for the measurement of excitonic resonances with linewidths as low as 8 meV (\cite{vaquero2020excitons}). Our results show that the $p$ resonance in the shift photocurrent  
splits up by 100 meV from the B-$1s$ exciton peak, and gives a comparable current response. It is thus of special importance to take into account the possible impact of this feature when interpreting photogalvanic measurements on TMDC-based experiments.

\section{Conclusion and perspective}
\label{sec:iv}

In this work, we have set up a methodology to calculate excitonic non-linear optical responses based on Wannierized band structures combined with effective electron-hole model interactions. To perform the calculations, we have interfaced the output of the \textsc{Wannier90} code with our recently-developed package \textsc{Xatu} for solving the BSE. We have developed a post-processing code to calculate the optical response in the linear and second-order regimes. 
This methodology updates previous works on tight-binding excitons, now by using accurate wannierization of ab initio band structures. This provides several advantages: multiband effects are captured, the band structure is obtained accurately (including effective masses), and inter-site matrix elements are correctly accounted for. 
This method thus offers a better description of linear and nonlinear optical responses than 
empirical tight-binding models.

As examples, we have addressed the role of in-gap excitons in the linear absorbance and shift conductivity of monolayers MoS$_2$ and GeS. At low energies, we have shown that the total short circuit current is enlarged by nearly one order of magnitude in GeS due to exciton effects, while this enhancement is lower in Mo$S_2$. Remarkably, we have shown that $2p$ excitons in both cases are clearly visible in the photocurrent spectrum of both materials. 

While experimental measurements on WS$_2$ do not display any measurable shift-current response, a clear photogalvanic current (under circularly-polarized light) is identified on Ref. \cite{quereda2018symmetry} in monolayer MoSe$_2$, showing excitonic features. No doubt, more experiments are needed to characterize the exciton fine-structure in the BPVE response in TMDC monolayer samples. 

By extending the electron-hole interaction kernel beyond two dimensions, a large variety of systems will be able to be addressed in future works. The post-processing tools used in this work will be available soon in a the next version of the \textsc{Xatu} code \cite{uria2023efficient}, via the package repository.

\begin{acknowledgments}
The authors acknowledge financial support from the Spanish MICINN (grants nos. PID2019-109539GB-C43, TED2021-131323B-I00, and PID2022-141712NB-C21), the Mar\'ia de Maeztu Program for Units of Excellence in R\&D (grant no. CEX2018-000805-M), Comunidad Aut\'onoma de Madrid through the Nanomag COST-CM Program (grant no. S2018/NMT-4321), Generalitat Valenciana through Programa Prometeo (2021/017), Centro de Computaci\'on Cient\'ifica of the Universidad Aut\'onoma de Madrid, and Red Española de Supercomputaci\'on. M.C.-G. is grateful to the Azrieli Foundation for the award of an Azrieli International Postdoctoral Fellowship. Additional computational resources were provided by the Weizmann Institute of Science at Chemfarm. M.C.-G. thanks Tonatiuh Rangel for providing the initial geometries of bulk and monolayer GeS.
\end{acknowledgments}

\bibliographystyle{unsrt}
\bibliography{biblio} 
\end{document}